\documentclass[prl,aps,twocolumn,showpacs,floatfix,10pt]{revtex4}
\usepackage[dvips]{graphicx}
\usepackage{amsmath}
\usepackage{amssymb}
\usepackage[nice]{nicefrac}
\usepackage{color}

\begin{document}

	\title{Scaled Brownian motion as a mean field model for continuous time random walks}

	\author{Felix Thiel}
	\email{thiel@physik.hu-berlin.de}
	\affiliation{Institut f{\"u}r Physik, Humboldt-Universit{\"a}t zu Berlin, Newtonstrasse 15, D--12489 Berlin, Germany}

	\author{Igor M. Sokolov}
	\email{igor.sokolov@physik.hu-berlin.de}
	\affiliation{Institut f{\"u}r Physik, Humboldt-Universit{\"a}t zu Berlin, Newtonstrasse 15, D--12489 Berlin, Germany}

	\begin{abstract}
	We consider scaled Brownian motion (sBm), a random process described by a diffusion equation with explicitly time-dependent diffusion
	coefficient $D(t) = D_0 t^{\alpha-1}$ (Batchelor's equation) which, for $\alpha < 1$, is often used for fitting experimental data for subdiffusion of unclear genesis. 
	We show that this process is a close relative of subdiffusive continuous-time random walks and describes the motion of the center of mass of a cloud of independent 
	walkers. It shares with subdiffusive CTRW its non-stationary and non-ergodic properties. The non-ergodicity of sBm does not however go hand in hand with strong
	difference between its different realizations: its heterogeneity (``ergodicity breaking'') parameter tends to zero for long trajectories. 	 
	\end{abstract}


	\date{\today}

	\maketitle

	Anomalous diffusion is a generic name for a class of transport processes which are close to diffusion in their origin (i.e. can be represented via generalized random 
	walk schemes or Langevin equations) but do not lead to the mean squared displacement growing as the first power of time 
	\begin{equation}
	 \langle \mathbf{r}^2(t) \rangle = 2dDt
	 \label{Diff}
	\end{equation}
	(with $D$ being the diffusion coefficient and $d$ the dimension of space), as predicted by the Fick's laws.
	Within the random walk schemes such deviations from the normal diffusion picture can arise either due to broad distributions of the waiting times between the steps 
	(continuous time random walk models, CTRW), or due to slow decay of correlations between steps, or both, see \cite{Sokolov} for a review, leading to the change of 
	the power law in the time dependence of the mean squared displacement, 
	\[
		\langle \mathbf{r}^2(t) \rangle \propto t^\alpha.
	\]
	The processes with $\alpha < 1$ are called subdiffusion, the ones with $\alpha > 1$ are termed superdiffusion.
	In the first case the formal diffusion coefficient $D$ in Eq.(\ref{Diff}) vanishes in the long time limit; in the second case it diverges.

	The single trajectory dynamics in normal diffusion is described by the Langevin equation 
	\[
	 \dot{x} = \sqrt{2D} \xi(t)
	\]
	with white, delta-correlated Gaussian noise $\xi(t)$, $\langle \xi(t) \rangle = 0$, $\langle \xi(t) \xi(t') \rangle = \delta(t-t')$; the time-dependence of the 
	probability density function (PDF) of the process or the one of its transition probabilities is given by the Fick's second law (diffusion equation) 
	\[
		\frac{\partial}{\partial t} p(x,t) = D \frac{\partial^2}{\partial x^2} p(x,t)
	\]
	(both equations given here in one dimension).
	The description of anomalous diffusion of different origins often follows by modification of one of the equations above.

	In experiments, many processes of anomalous diffusion of unknown origin, i.e. when the observable of interest  which cannot be fitted to the solutions of Eq.(\ref{Diff}), 
	are fitted to the results obtained for the so-called scaled Brownian motion (sBm) \cite{Lim}, a diffusion process with explicitly time-dependent diffusion coefficient $D(t) = D_0 t^{\alpha-1}$.
	Numerical simulations of Ref. \cite{Saxton} show that, at least for the case of FRAP (fluorescence recovery after photobleaching), the fits may be astonishingly good,
	independent on the true nature of the simulated process (percolation, CTRW, etc.). This means that the form of FRAP recovery curves, if they hint onto
	anomalous diffusion, hardly depends on the origin of the corresponding anomaly. Using sBm model for calculating other properties may however be dangerous, as long as
	the nature of the sBm model itself and the one of the process under investigation are not well-understood \cite{Sokolov}.
	
	The PDF and the transition probabilities in sBm are given by the Batchelor's equation \cite{Batchelor},
	\begin{equation}
	 \frac{\partial}{\partial t} p(x,t) = D_0 t^{\alpha -1} \frac{\partial^2}{\partial x^2} p(x,t),
	 \label{Batchelor}
	\end{equation}
	leading to the mean squared displacement $\langle x^2(t) \rangle \propto t^\alpha$ (the original one was for $\alpha = 3$).
	Initially, the Eq.(\ref{Batchelor}) was proposed for description of superdiffusive turbulent dispersion, as an alternative to the Richardson's diffusion 
	equation with the distance-dependent diffusion coefficient. Some of its shortcomings for description of the turbulent dispersion were clear to Batchelor himself,
	see Ref. \cite{Monin} for more detailed discussion. The equations of Batchelor's type are typically postulated and do not follow from any explicit physical model, 
	one of seldom exclusions being \cite{Postnikov}.
	
	The Langevin description of the corresponding process is given by 
	\begin{equation}
			\dot{x} 
		= 
			\sqrt{2 D_0 \alpha t^{\alpha -1}} \xi(t)
		\label{Langevin}
	\end{equation}
	with white, delta-correlated Gaussian noise $\xi(t)$, $\langle \xi(t) \rangle = 0$, $\langle \xi(t) \xi(t') \rangle = \delta(t-t')$.
	Both, the Langevin and the diffusion equation for sBm can be reduced to the ones for the normal one by the time rescaling $t \to \tau = t^\alpha$.
	Thus, sBm is a random process which is subordinated to a Brownian motion (Wiener process) under the deterministic time change given above 
	and as such is a relative of the CTRW with the only (but important) difference that in CTRW the time change is stochastic.
	The qualitative discussion of the relation between the sBm and the CTRW was given in \cite{Sokolov} without proofs and calculations. Here we close this
	gap and provide the deeper analysis of sBm. Thus, we show that sBm can be considered as a homogenized (mean field) 
	approximation to CTRW and shares its property of aging and ergodicity breaking, but this ergodicity breaking does not go hand in hand with inhomogeneity 
	(non-convergence in distribution of the mean squared displacement as obtained by the moving time average). This stresses that the absence of ergodicity 
	(due to non-stationarity) does not imply the non-zero value of the ``ergodicity breaking parameter'' which characterizes the heterogeneity of realizations.
	In sBm, although a close relative of CTRW, this heterogeneity is removed by the pre-averaging procedure.
	In what follows we concentrate on the subdiffusive case $\alpha < 1$, as considered in \cite{Saxton}.

	Let us first discuss a general situation and consider a random process $x_i(t)$ in continuous time $t$.
	Let us assume that the process possesses all necessary single-point and cross-moments.
	We will associate this random process with the coordinate of the $i$-th walker at time $t$.
	The process will be taken to possess zero mean.
	Let us now consider the behavior of the center of mass of $m$ independent walkers (i.e. the mean coordinate of $m$ independent random processes 
	$x_i(t)$, $i=1,2, ... , m$) $X(t) = (1/m) \sum_{i=1}^m x_i(t)$.
	Note that $\langle x_i(t) \rangle = 0$ due to the symmetry of the process, $\langle x_i(t) x_j(t) \rangle = 0$ (for $i \neq j$ ) due to independence 
	of different realizations, and $\langle x_i(t) x_i(t) \rangle = \sigma(t)$, so that $\langle X^2 (t) \rangle = \sigma^2(t) / m$.
	In order to prove the corresponding limit theorems we need to define the mean position in such a way that its second moment does not depend on 
	$m$ i.e. to rescale $x^{(m)}(t) \to X(t) \sqrt{m}$, or, in other words, to redefine the ``center of mass'' position as 
	 \[
	  x^{(m)} (t) = \frac{1}{\sqrt{m}} \sum_{i=1}^m x_i(t).
	 \]
	Let us note that the random process $x^{(m)} (t)$ giving the rescaled center of mass  position retains the correlation function of the initial $x_i(t)$ process.
	Note that for symmetric processes with correlation function $\langle x_i(t) x_j(t') \rangle = C(t,t')\delta_{ij}$, $x^{(m)}(t)$ has the same correlation function:
	 $ \langle x^{(m)}(t) x^{(m)}(t') \rangle = \frac{1}{m} \sum_{i,j=1}^m \langle x_i(t) x_j(t') \rangle 
	 = \frac{1}{m} \sum_{i,j=1}^m C(t,t') \delta_{ij} = \frac{1}{m} \sum_{i=1}^m C(t,t') = C(t,t')$.
	 The mean squared change in $x^{(m)}$, given by $C(t,t)$, does not depend on $m$.
	 In the limiting case, $x(t) = \lim_{m\rightarrow\infty} x^{(m)}(t)$ will correspond to the ``mean field coordinate'' (MF-position) of the walker.
	The position of the center of mass of the cloud is obtained by inverse rescaling.

	We now show that the position $x(t)$ and all possible vectors $(x(t_1), x(t_2),..., x(t_n))$ comprising the walker’s positions at different times converge to Gaussian, 
	so that the process $x^{(m)}(t)$ converges to a Gaussian process for $m \to \infty$.
	
	Let us consider a sequence of times $t_1$,$t_2$,...,$t_n$ and a corresponding vector $\mathbf{X}_i = (x_i(t_1), x_i(t_2),..., x_i(t_n))$.
	This vector (stemming from probing the position of the $i$-th walker at time instants $t_1,t_2, ..., t_n$) does possess a mean $\langle \mathbf{X}_i \rangle = 0$ 
	and a covariation matrix between its components with finite elements $C_{jk}=C(t_j,t_k)$.
	According to the central limit theorem for multivariate (vector) distributions the corresponding mean  $\mathbf{X}^{(m)} = \frac{1}{\sqrt{m}} \sum_{i=1}^m \mathbf{X}_i$ 
	converges in distribution to a multivariate Gaussian with zero mean and the same covariation matrix, see e.g. \cite{Feller,Mukhopadhyay}.
	Since this happens for any sequence of observation times, the limit process, the one describing the ``center-of-mass motion'' of a cloud of walkers, is a Gaussian process
	with the correlation function inherited from the single realization $x_i(t)$.
	This process can be considered as a kind of ``mean-field approximation'' for our initial process $x_i(t)$.
	
	We have seen that pooling (superimposing many statistical copies of the initial process) leads us to a Gaussian process with the correlation function inherited from a single copy.
	Now we turn to $x_i(t)$ being a CTRW with the power-law waiting time density $\psi(t) \simeq \tau^\alpha t^{-1 - \alpha}$
	and the mean squared displacement $\sigma^2(t) \simeq 2 D_0 t^\alpha$ with $D_0$ being the combination of mean squared displacement per step $a^2$ and typical waiting time $\tau$ \cite{Klafter}.
	Let us subdivide the time axis in short intervals of duration $d t$, and resample each of the CTRW processes as a simple random walk with the step duration $dt$ 
	and with three possible step lengths $s_i$ of zero and $\pm 1$ (with $i = t/dt$; the double steps during the $dt$ intervals can be neglected provided $dt$ is small enough).
	The steps of this process are not independent (if the leading process is not a Poissonian one, i.e. if the waiting time distributions are not exponential), but always uncorrelated, 
	because of the symmetry: $\langle s_i s_j \rangle =0$ for $i \neq j$ i.e. for the resampled process $\langle s_i s_j \rangle =\delta_{ij}$.
	The displacements of a walker during two non-intersecting time intervals $\Delta t_1 = t_2 - t_1$, $\Delta X_1 = \sum_{i=[t_1/dt]}^{[t_2/dt]} s_i$,  
	and $\Delta t_2 = t_4 - t_3$, $\Delta X_2 = \sum_{j=[t_3/dt]}^{[t_4/dt]} s_j$ (the square brackets here denote the whole part of the corresponding number) are non-correlated since, 
	for all possible $i$ and $j$, $i \neq j$.
	Thus, for two non-intersecting time intervals $\langle \Delta X_1 \Delta X_2 \rangle =0$.
	This observation allows us to get the position-position correlation function $C(t,t')= \langle x(t) x(t') \rangle$ in CTRW.
	Taking $t' > t$ on can put $C(t,t') = \langle x(t) [x(t) + \Delta x(t'-t) \rangle = \langle x^2(t) \rangle + \langle \Delta x(t' -t) x(t) \rangle$ with $\Delta x(t'-t)$ being 
	the walker's displacement during the time interval of duration $t'-t$ starting at $t$.
	The mean $\langle \Delta x(t' -t) x(t) \rangle$ vanishes as discussed above.
	Therefore the correlation functions $C(t,t')$ for CTRW are given by 
	\begin{equation}
	 C(t,t') = \langle x^2( \min(t,t') ) \rangle = 2 D_0 [\min(t,t')]^{\alpha}.
	 \label{CF}
	\end{equation}

	We now consider the superposition of $m$ independent CTRWs.
	When the number $m$ of independent pooled (superimposed) random processes tends to infinity, so does also the number of events within each interval of a fixed 
	length, and the displacement $\Delta x$ of a pooled process during this interval tends to a Gaussian.
	The displacements at different $\Delta t$-intervals are non-correlated Gaussian random variables and are therefore independent \cite{Feller}: 
	the dependence present in the values of $x_i(t)$ gets ``dissolved'' when the number of processes tends to infinity.
	The mean squared displacement during the $\Delta t$-interval starting at $t$ is given by 
	\begin{equation}
	\langle \Delta x^2 \rangle= 2D_0 (t + \Delta t)^\alpha - 2D_0 t^\alpha \simeq 2 D_0 \alpha t^{\alpha -1} \Delta t.
	\label{displacement}
	\end{equation}
	We then can write $x(t+\Delta t) = x(t) + \Delta x(t)$ and interpret it as a finite-difference approximation to a Langevin equation 
	\[
	 \frac{d x}{d t} = \eta(t),
	\]
	i.e. as the result of integrating this equation over the finite time interval $\Delta t$ starting at $t$. Since the distribution of $\Delta x(t)$
	is Gaussian, the values of $\Delta x$ at different time intervals are uncorrelated, and $\langle \Delta x^2 \rangle = 2 D_0 \alpha t^{\alpha -1} \Delta t$,
	the properties of the noise follow: this noise is Gaussian with $\langle \eta(t) \rangle = 0$ and $\langle \eta(t) \eta(t')\rangle =2 D_0 \alpha t^{\alpha -1} \delta(t-t')$,
	i.e. exactly as in Eq.(\ref{Langevin}). 
	This diffusion process with explicitly time-dependent diffusion coefficient is the sBm, and the Batchelor's equation, Eq.(\ref{Batchelor}) appears as a 
	Fokker-Planck equation for the Langevin equation above.
	The mean field process shares with the CTRW the same time-dependence of the mean squared displacement; $\langle x^2(t) \rangle \propto t^\alpha$ in the ensemble average.
	Therefore, the double, ensemble and moving time average, goes as 
	\begin{eqnarray*}
	&& \langle \overline{x^2(t)} \rangle = \frac{1}{T-t} \int_0^{T-t} \langle [x(t + \tau) - x(\tau)]^2 \rangle d \tau \\
	 && =  2 D_0  \frac{1}{T-t} \int_0^{T-t} [(t +\tau)^\alpha - \tau^\alpha] d\tau \approx 2 D_0 T^{\alpha - 1} t
	\end{eqnarray*}
	for $t \ll T$, just like in CTRW \cite{Lubelski}, showing ergodicity breaking (a trivial one, due to the non-stationarity of the process).
	At difference with CTRW, the ``ergodicity breaking parameter'' for this process vanishes, showing that its different realizations are extremely similar.

	Let $\overline{x^2(t)} = \frac{1}{T-t} \int_0^{T-t} \left[ x(\tau + t) - x(\tau) \right]^2 d \tau$
	be the moving time averaged mean squared displacement.
	In CTRW this one is a random variable: its distribution does not narrow for $T\rightarrow$ in the sense that the heterogeneity (``ergodicity breaking'') parameter $EB$ \cite{Barkai}
	\begin{equation}
		EB = \frac{ \left\langle \overline{x^2(t)}^2 \right\rangle - \langle \overline{x^2(t)} \rangle^2 }{ \langle \overline{x^2(t)} \rangle^2 }
		\label{EBeq}
	\end{equation}
	tends to a finite limit $EB = 2 \Gamma^2(1+\alpha)/ \Gamma(1+2\alpha) -1$ for $T\rightarrow\infty$.
	Expressing $EB$ via correlation functions we see that vanishing of this parameter for a \textit{stationary} process leads to ergodicity (i.e. equality of the ensemble mean and
	the moving time average) as a consequence of Birkhoff-Khinchin theorem, stressing that ergodicity implies stationarity and some amount of ``homogeneity''.
	Scaled Brownian motion (Batchelor's process) is non-stationary (since it is explicitly time-inhomogeneous), and non-ergodic, but shows high amount of homogeneity among its trajectories, 
	as we proceed to show by explicit calculation of $EB$ and showing that it vanishes for $T \rightarrow \infty$.
	
	The only thing we need to calculate is $\langle \overline{x^2(t)}^2\rangle$:
	\begin{eqnarray}
		&& \left\langle \overline{x^2(t)}^2 \right\rangle
		=  \frac{1}{(T-t)^2} \left\langle \int_0^{T-t}d \tau_1 \left[ x(\tau_1 + t) - x(\tau_1) \right]^2 \right. \nonumber \\
		&& \qquad \times \left. \int_0^{T-t}d \tau_2 \left[ x(\tau_2 + t) - x(\tau_2) \right]^2 \right\rangle \nonumber \\
		&& =  \frac{2}{(T-t)^2} \left\langle \int_0^{T-t}d \tau_d \int_{0}^{T-t-\tau_d} d \tau_m  
			\left[ x(\tau_m + t) \right. \right. \nonumber \\
		&& \qquad \left. - x(\tau_m) \right]^2  \left[ x(\tau_m + \tau_d + t) - x(\tau_m + \tau_d) \right]^2  \bigg\rangle,
		\label{LongInt}
	\end{eqnarray}
	where we changed the variables of integration to $\tau_m = \min(\tau_1,\tau_2)$, $\tau_d = |\tau_2 - \tau_1|$ and used the symmetry of the expression.
	The order of integration and ensemble averaging can be interchanged and the mathematical expectation is readily evaluated by using the Gaussian property of $x(t)$.
	Let us denote by $\tilde{C}(\tau_1,\tau_2) = \langle [x(\tau_1+t) - x(\tau_1)][ x(\tau_2+t) - x(\tau_2)] \rangle$ the correlation function of the increments.
	Expressed in variables of Eq.(\ref{LongInt}) $\tilde{C}$ reads: $\tilde{C}(\tau_m,\tau_m+\tau_d) =  2 D_0 [(\tau_m+t)^{\alpha} - (\min(\tau_m+t,\tau_m+\tau_d))^{\alpha}] 								
	=  2 D_0 [ (\tau_m+t)^{\alpha} - (\tau_m+\tau_d)^{\alpha}] I_{[0,t)}(\tau_d)$, 
	where $I_A(x)$ is an indicator function that equals unity if $x \in A$ and vanishes otherwise. 
	Note, that the increment process is also Gaussian, so that its higher correlators decompose into products of $\tilde{C}(\tau_1,\tau_2)$.
	Thus the integrand in the last expression of Eq.(\ref{LongInt}) reads:
	\begin{align*}
		&
			\left\langle 
				\left[ x(\tau_m + t) - x(\tau_m) \right]^2 
				\left[ x(\tau_m + \tau_d + t) - x(\tau_m + \tau_d) \right]^2
			\right\rangle												\\
		& = 
			2 \tilde{C}^2(\tau_m,\tau_m+\tau_d) + \tilde{C}(\tau_m,\tau_m) \tilde{C}(\tau_m+\tau_d,\tau_m+\tau_d) 		\\
		& = 
			8 D_{0}^{2} [ (\tau_m+t)^{\alpha} - (\tau_m+\tau_d)^{\alpha}]^2 I_{[0,t)}(\tau_d) 					\\
		& \quad
			+ \tilde{C}(\tau_m,\tau_m) \tilde{C}(\tau_m+\tau_d,\tau_m+\tau_d)					
	\end{align*}
	The last summand can be identified with the $\langle \overline{x^2(t)}\rangle^2$ term in the numerator of Eq. (\ref{EBeq}) and therefore cancels out.
	The remaining summand vanishes, if $\tau_d > t$.
	This is the main difference to usual CTRW, where the existence of such longer correlations is responsible for non-vanishing EB-parameter.

	Now let us assume $T$ to be sufficiently large, i.e. $T~-~t~\gg~t$, and proceed.
	\begin{align}
		\nonumber &
			\left\langle \overline{x^2(t)}^2 \right\rangle - \langle \overline{x^2(t)} \rangle^2	
		= 
			\frac{ 16 D_{0}^{2} }{ (T-t)^2 }  \\
		&	\times \int_{0}^{t} d \tau_d \int_{0}^{T-t-\tau_d} d \tau_m
			[ (\tau_m+t)^{\alpha} - (\tau_m+\tau_d)^{\alpha}]^2	\text{.}
			\label{Nasty}
	\end{align}
	For large $\tau_m$ the integrand is asymptotically equal to $\alpha^2 \tau_{m}^{2\alpha-2} (t-\tau_d)^2$.
	For $\alpha > 1/2$ this asymptotic form can be immediately plugged into Eq. (\ref{Nasty}) and the corresponding integral 
	readily evaluated. Its limit for $T \gg t$ is:
	\[
	\left\langle \overline{x^2(t)}^2 \right\rangle - \langle \overline{x^2(t)} \rangle^2
		= \frac{ 16 \alpha^2 D_{0}^{2} t^3 }{3 (2\alpha - 1) (T-t)^{3-2\alpha} } .
	\]
	Inserting the corresponding expression
	into Eq.(\ref{EBeq}) shows that $EB$ approaches zero as $T^{-1}$.

	For $\alpha \leq 1/2$ the substitution of the integrand by its asymptotic expression for large $\tau_m$
	leads to the divergence of the inner integral in Eq.(\ref{Nasty}) at the lower limit of integration.  
	Let us split the integral into two parts at some intermediate time $t_{i} > 0$ and use the asymptotical substitution discussed above only in the second part: 
	$\int_{0}^{t} d \tau_d \int_{0}^{T-t-\tau_d} d \tau_m 
	[ (\tau_m+t)^{\alpha} - (\tau_m+\tau_d)^{\alpha}]^2 = \int_{0}^{t_{i}} d \tau_m [ (\tau_m+t)^{\alpha} - (\tau_m+\tau_d)^{\alpha}]^2 + 
	\alpha^2 \int_{t_{i}}^{T-t-\tau_d} d \tau_m (t - \tau_d)^2 \tau_m^{2\alpha-2}$. 
	The first summand is bounded from above by $t_i [t^{\alpha} - \tau_d^{\alpha}]^2$, and the second one converges on the upper limit, which thus
	can be taken to go to infinity. Therefore the inner integral (and thus the whole double integral in Eq.(\ref{Nasty})) tends for $T$ large to a function independent on $T$.
	We thus conclude that
	\begin{equation}
			\frac{ \left\langle \overline{x^2(t)}^2 \right\rangle - \langle \overline{x^2(t)} \rangle^2 }{ \langle \overline{x^2(t)} \rangle^2 }
		\simeq 
			\left\{ \begin{aligned}
				4 Z_{\alpha} \left( \frac{t}{T} \right)^{2\alpha} \text{,} \qquad \alpha \le \frac{1}{2}				\\
				\frac{4 \alpha^2}{ 3 (2\alpha-1)} \frac{t}{T} \text{,} \qquad \alpha > \frac{1}{2}
			\end{aligned} \right.
		\label{EBResult}
	\end{equation}
	with $Z_{\alpha} = \int_{0}^{1} dy \int_{0}^{\infty} dx [ (x+1)^{\alpha} - (x+y)^{\alpha}]^2$. 

	We see, that for any positive value of $\alpha$ $EB$ will vanish, and shows a crossover between two types of $T$-dependence at $\alpha=1/2$.  
	Figure~\ref{Pic} shows simulated values of $EB$ for different values of $\alpha$ and $T$ in a double logarithmic plot together with the predictions of Eq(\ref{EBResult}).

	In conclusion: Scaled Brownian motion (Batchelor's process), described by a diffusion equation with explicitly time-dependent diffusion coefficient and 
	often used in fitting data for experimental situations showing anomalous diffusion of unclear origin, is, in subdiffusive case $\alpha <1$, a close relative of CTRW and describes
	the motion of the center of mass of a cloud independent continuous time random walkers. The model shows the same kind of ergodicity breaking due to
	non-stationarity as the corresponding CTRW process. This one, however, does not go hand in hand with strong heterogeneity of its different realizations:
	the corresponding ``ergodicity breaking'' parameter vanishes in the limit of long trajectories. 
	
	\acknowledgements{The authors acknowledge financial support by DFG within IRTG 1740 research and training group project.

	\begin{figure}
		\includegraphics[width=0.5\textwidth]{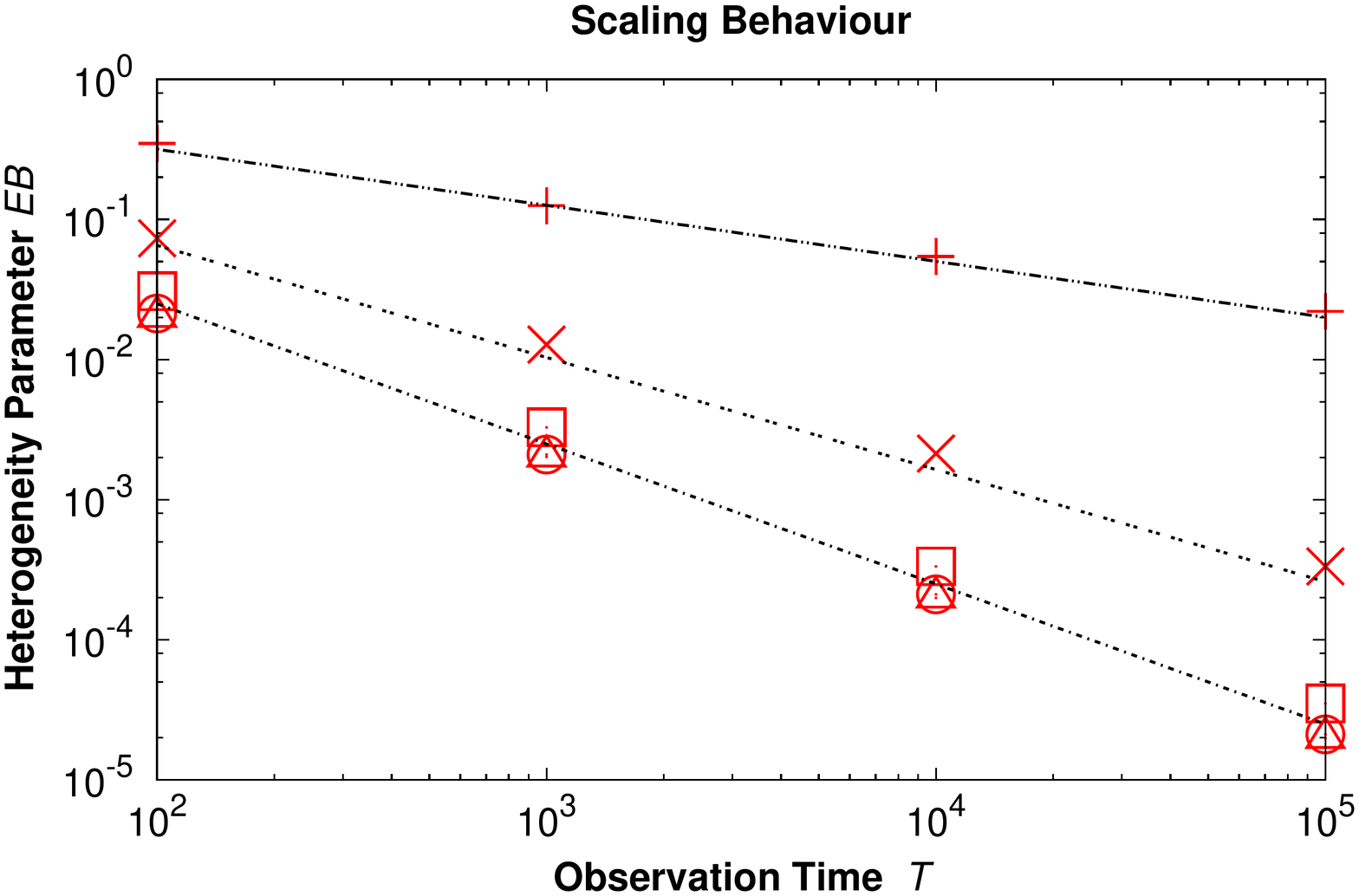}
		\caption{
			Scaling Behaviour of $EB$.
			The Batchelor's Process was simulated and $EB$ was calculated for different values of $\alpha$ and $T$.
			$\alpha$-values range from $0.2$ ($+$) , $0.4$ ($\times$), $0.6$ ($\Box$), $0.8$ ($\odot$), and $1.0$ ($\triangle$).
			Each point is calculated from $5000$ trajectories. The given power laws are (from upper to lower line ) $T^{-0.4}$, $T^{-0.8}$, and $T^{-1}$.
			\label{Pic}
		}
	\end{figure}


\begin{thebibliography}{99}
		\bibitem{Sokolov}  I.M. Sokolov, Soft Matter, \textbf{8}, 9043 (2012)
		\bibitem{Lim} S. C. Lim and S. V. Muniandy, Phys. Rev. E \textbf{66}, 021114 (2002)
		\bibitem{Saxton} M. J. Saxton, Biophys. J. \textbf{81}, 2226 (2001)
		\bibitem{Batchelor} G.K. Batchelor, Math. Proc. Cambridge Phil. Soc., \textbf{48}, 345 (1952) 
		\bibitem{Monin} A.S. Monin, A.M. Yaglom, \textit{Statistical Fluid Mechanics: Mechanics of Turbulence}, Volume 1, M.I.T. Press, Cambridge, Mass. (1971)	
		\bibitem{Postnikov} E.B. Postnikov and I.M. Sokolov, Physica A \textbf{391}, 5095 - 5101 (2012)
		\bibitem{Feller} W. Feller, \textit{An Introduction to Probability Theory and Its Applications}, Wiley, ... Vol. II Chapter III.
		\bibitem{Mukhopadhyay} P. Mukhopadhyay, \textit{Multivariate Statistical Analysis}, World Scientific, Singapore, 2009 (see Sec. 3.5).
		\bibitem{Klafter} J. Klafter and I.M. Sokolov, \textit{First Steps in Random Walks: From Tools to Applications}, Oxford university Press (2011)
		\bibitem{Lubelski} A. Lubelski, I.M. Sokolov and J. Klafter, Phys. Rev. Lett. \textbf{100}, 250602 (2008)
		\bibitem{Barkai} Y. He, S. Burov, R. Metzler, and E. Barkai, Phys. Rev. Lett. \textbf{101}, 058101 (2008)
	\end{thebibliography}
\end{document}